\documentclass[reprint, superscriptaddress, secnumarabic, amssymb, nobibnotes, aps, prl]{revtex4-1}

\usepackage{graphicx}
\usepackage{epstopdf}
\usepackage[T1]{fontenc}
\usepackage[latin9]{inputenc}
\usepackage{amsbsy}
\usepackage{gensymb}

\begin{document}
\title{\textrm{Time Reversal Symmetry Breaking in noncentrosymmetric superconductor Re$_{6}$Ti}}
\author{D. Singh}
\affiliation{Indian Institute of Science Education and Research Bhopal, Bhopal, 462066, India}
\author{Sajilesh K.P}
\affiliation{Indian Institute of Science Education and Research Bhopal, Bhopal, 462066, India}
\author{J. A. T. Barker}
\affiliation{Physics Department, University of Warwick, Coventry CV4 7AL, United Kingdom}
\author{D. McK. Paul}
\affiliation{Physics Department, University of Warwick, Coventry CV4 7AL, United Kingdom}
\author{A. D. Hillier}
\affiliation{ISIS facility, STFC Rutherford Appleton Laboratory, Harwell Science and Innovation Campus, Oxfordshire, OX11 0QX, UK}
\author{R. P. Singh}
\email[]{rpsingh@iiserb.ac.in}
\affiliation{Indian Institute of Science Education and Research Bhopal, Bhopal, 462066, India}

\date{\today}
\begin{abstract}
\begin{flushleft}

\end{flushleft}
We have investigated the superconducting state of the noncentrosymmetric superconductor Re$_{6}$Ti (T$_{c}$ = 6.0 K) using muon-spin rotation/relaxation ($\mu$SR) technique. The zero-field muon experiment shows the presence of spontaneous magnetic fields in the superconducting state, indicating time-reversal symmetry breaking (TRSB). However, the low-temperature transverse field muon measurements suggest nodeless s-wave superconductivity. The time reversal symmetry breaking further confirmed in the stoichiometric composition Re$_{24}$Ti$_{5}$. These results indicate that the pairing symmetry is not affected by spin-orbital coupling Re$_{6}$X family of compounds. Altogether these studies suggest unconventional nature (TRSB) of superconductivity is intrinsic to Re$_{6}$X family of compounds and paves the way for further studies of this family of materials.
\end{abstract}
\maketitle
Superconductors with noncentrosymmetric crystal structure are of considerable interest due to their possible realization of unconventional superconductivity \cite{MS, MSS1}. The lack of inversion symmetry in the lattices of these materials has significant implications on the symmetry of their superconducting state. The direct consequence of the broken inversion symmetry was first recognized in the noncentrosymmetric superconductor (NCS) CePt$_{3}$Si \cite{EBG}. It shows upper critical field exceeding Pauli limiting field, indicating unconventional behaviour \cite{EBG,EBM,EBH}. In an NCS, the lack of inversion symmetry introduces a Rashba-type antisymmetric spin-orbit coupling (ASOC) \cite{EIR,LPG}, which results in the splitting of spin-up and spin-down conduction electron energy bands. This allows the mixing of orbital and spin parts of the Cooper wave function, which leads to parity mixed superconductivity. The extent of the parity mixing is determined by the strength of the ASOC. The rigorous experimental and theoretical investigations on NCSs have been carried out in recent time to understand their complex superconductivity properties. Indeed, unconventional superconductivity found in several noncentrosymmetric superconductors. The examples are Li$_{2}$(Pd,Pt)$_{3}$B \cite{HQY,MNY,HTM,SHJ}, Mo$_{3}$Al$_{2}$C \cite{ABK} Re$_{3}$W \cite{PKB}, Nb$_{0.18}$Re$_{0.82}$ \cite{ABKY,CSL}, Y$_{2}$C$_{3}$ \cite{JCM} etc.\\

Recently, time-reversal symmetry breaking (rarely observed phenomenon) has been observed in few unconventional superconductors \cite{SR1,SR2,UP1,UP2,PPG,LNG,LRS}. Due to parity mixed superconductivity, NCS are prime candidates to exhibit this rarely observed phenomenon. To date, it has been reported to be observed  only few NCS materials, LaNiC$_{2}$ \cite{ADH}, Re$_{6}$X (X= Zr,Hf) \cite{RPS,DSJ}, locally noncentrosymmetric SrPtAs \cite{PKB2} and La$_{7}$Ir$_{3}$ \cite{JAT}. On the other hand, it found to be preserved in several other NCSs \cite{ABK,PKB,ABKY,JCM,GED,VKA,TKF}.

Our recent work primarily focused on the role of ASOC in controlling the parity mixing in NCS materials. The systems containing heavier transition metal elements are of particular interest since there, the strength of spin-triplet component in the pairing mixing ratio can be enhanced markedly. In this regard, we systemically began the investigation on Re based compounds with $\alpha$-Mn structure. The unconventional superconductivity was observed in few members of Re$_{6}$X family, e.g. Re$_{6}$Zr and Re$_{6}$Hf provides the evidence of TRSB in superconducting ground state \cite{RPS, DSJ}. The TRSB signal strength of both the compounds are same. This precludes the possibility of ASOC having a role in superconductivity of NCSs. Same time contradictory results were also reported in the nuclear quadrupole resonance (NQR) studies of the Re-Zr system, displaying the exponential decrease of 1/T$_{1}$ Hebel-Slichter peaks below T$_{c}$ \cite{RZ2}. Hence, the relation between the breaking of inversion symmetry and time-reversal symmetry is yet not resolved.\\

Following the line of investigation, we are studying another member of this family Re$_{6}$Ti. The stoichiometric compound Re$_{24}$Ti$_{5}$ \cite{RT} is already studied where the bulk measurements doesn't show any unconventional property. The electronic structure of Re$_{24}$(Nb,Ti)$_{5}$ studied from the first principle, shows the density of states (DOS) at the Fermi level being dominated by the Re-5d electrons. This could one of the reason for the similar behavior for the Re$_{6}$Zr and Re$_{6}$Hf. To investigate the effect of ASOC and the influence of the dominance of Re at the DOS, we have done the microscopic study of the Re$_{6}$Ti and Re$_{24}$Ti$_{5}$. In this work, we show that the time reversal symmetry is broken in the superconducting state of the binary transition metal compound Re$_{6}$Ti and  Re$_{24}$Ti$_{5}$. The TRSB signal in Re$_{6}$Ti and  Re$_{24}$Ti$_{5}$ is very similar to the signal observed in other compounds (Re$_{6}$Zr/Hf) of this family. These results indicate that the superconducting ground state is not affected by the spin-orbital coupling.

\begin{figure}
\includegraphics[width=1.0\columnwidth]{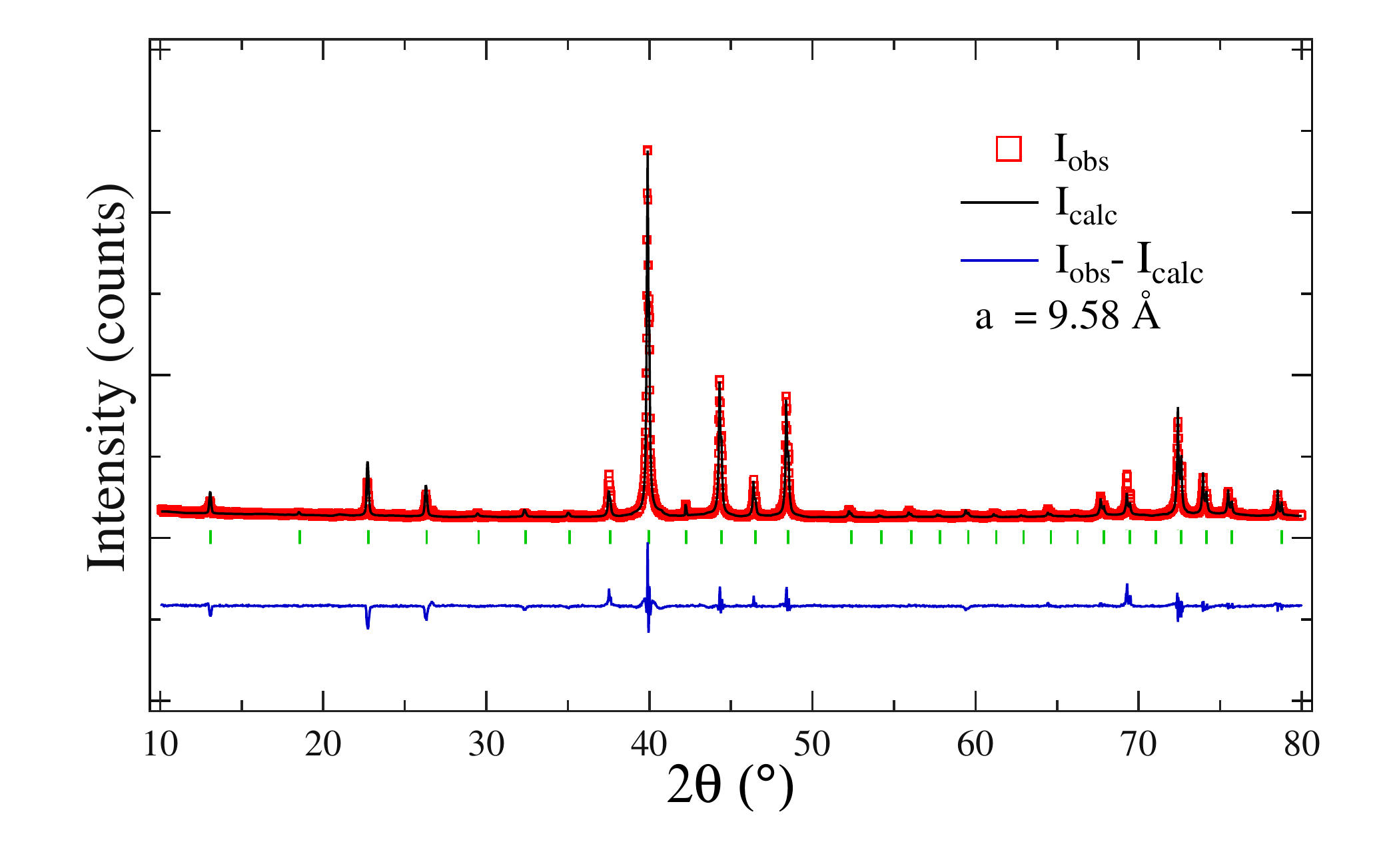}
\caption{\label{fig1:xrd} (color online) Powder XRD pattern for Re$_{6}$Ti sample recorded at room temperature using Cu $K_{\alpha}$ radiation. Rietveld refined calculated pattern for cubic non-centrosymmetric $\alpha$-$\textit{Mn}$ (217) structure shown by solid black line.}
\end{figure}

Polycrystalline samples of Re$_{6}$Ti were prepared by melting stoichiometric amounts Re (99.95$\%$, Alfa Aesar) and Ti (99.95$\%$, Alfa Aesar) in an arc melting furnace under ultrapure argon gas atmosphere on a water cooled copper hearth. The samples were inverted and remelted several times to ensure homogeneous mixing of constituent elements. The resulting sample was then sealed inside an evacuated quartz tube and annealed at 850 $^{\circ}$C for one week to remove any thermal stresses. The powder x-ray diffraction pattern for Re$_{6}$Ti was collected at room temperature. As observed from Fig. 1, the Re$_{6}$Ti sample has no impurity phase. It can be indexed by cubic, noncentrosymmetric $\alpha$ - $Mn$ structure (space group $I \bar{4}3m$, No. 217) with the lattice cell parameter a = 9.58(2) \text{\AA}.\\
The samples were characterized using magnetization and specific heat measurements. The appearance of strong diamagnetic signal at $T_{c}$ = 6.0 $\pm$ 0.02 K [Fig. 2 (a)], confirms superconducting transition in magnetization measurement. The low temperature specific heat measurement also confirms bulk superconductivity at $T_{c}$, where the normalized specific heat jump $\Delta C_{el}/\gamma_{n}T_{c}$ = 1.58 $\pm$ 0.02. This value is higher than the value for a BCS superconductor (1.43), suggesting strong coupling superconductivity. The results paralleled by the specific heat data in the superconducting state, where it fits perfectly well for a single-gap s-wave superconductor [Fig. 2(b)], for $\Delta (0)/k_{B}T_{c}$ = 1.86 (BCS value= 1.764).\\ 

\begin{figure}
\includegraphics[width=1.0\columnwidth]{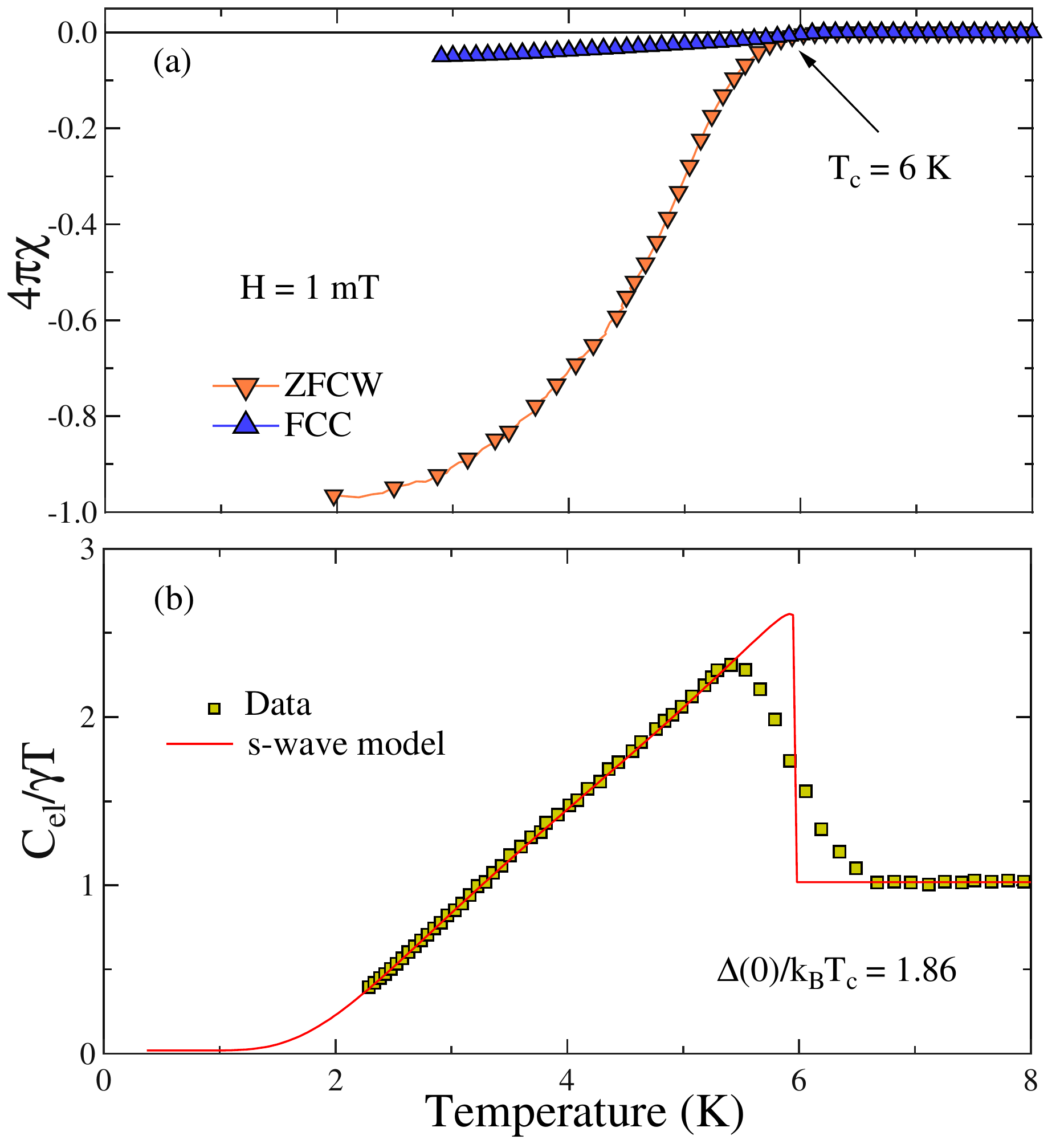}
\caption{\label{Fig2:Mt} (color online)(a) The superconducting transition temperature was appeared around $T_{c}$ = 6.0 $\pm$ 0.2 from magnetization measurement. (b) The low-temperature specific heat data in the superconducting state was fitted with single-gap s-wave model for $\Delta (0)/k_{B}T_{c}$ = 1.86.}
\end{figure}
To probe the superconducting ground state of Re$_{6}$Ti, muon spin relaxation and rotation measurements were carried out in the MuSR instrument at the ISIS pulsed muon and neutron spallation source.  A detailed account of the $\mu$SR technique may be found in Ref. \cite{SLL}. Stray fields at the sample position due to neighbouring instruments and the Earth's magnetic field are cancelled to within $\sim$1.0 $\mu$T using three sets of orthogonal coils and an active compensation system. The powdered Re$_{6}$Ti sample was mounted on a silver holder and placed in a sorption cryostat, which operated in the temperature range 0.3 K - 10 K.
\begin{figure}
\includegraphics[width=1.0\columnwidth]{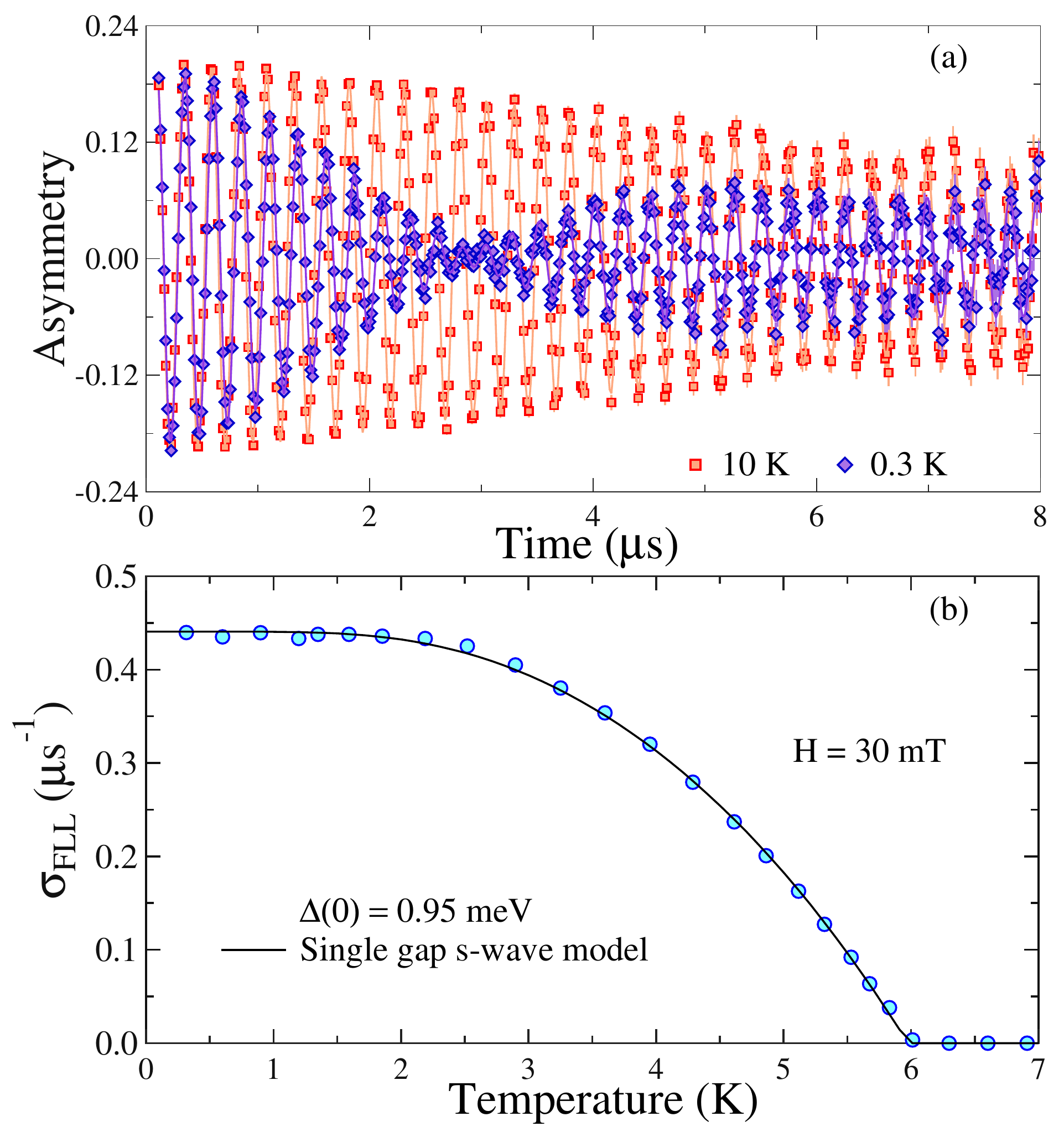}
\caption{\label{Fig3:TF} (color online) (a) Representative TF $\mu$SR signals collected at (a) 10 K and (b) 0.3 K in an applied magnetic field of 30 mT. The solid lines are fits using Eq. (1). (b) (a) Temperature dependence of $\sigma_{\mathrm{FLL}}$ collected at an applied field of 30 mT was following single gap s-wave model in dirty limit for $\Delta(0)/k_{B}T_{c}$ = 1.85 $\pm$ 0.01.}
\end{figure}
 Firstly, we performed transverse-field muon spin rotation (TF-$\mu$SR) measurements to directly measure the field distribution associated with the mixed state of a type-II superconductor to gain knowledge about the symmetry of pairing state. The sample was field-cooled in an applied field of 30 mT well above the lower critical field ($H_{c1}$(0) = 5.8 mT) in order to develop a well ordered flux line lattice (FLL). Asymmetry spectra were recorded above (10 K) and below (0.3 K) the transition temperature $T_{c}$ as displayed in Fig. 3 (a). In the normal state, the field distribution is homogeneous throughout the sample, which is depicted from the spectra taken at 10 K. The weak depolarization is attributed to the Gaussian relaxation that is due to the random nuclear dipolar field. The depolarization rate in the superconducting state becomes more prominent, due to the formation of inhomogeneous field distribution in the FLL state.\\
Time evolution of the asymmetry is best described by the sinusoidal oscillatory function damped with a Gaussian relaxation and an oscillatory background term arising from the muons implanted directly into the silver sample holder that does not depolarize,
\begin{equation}
G_{\mathrm{TF}}(t) = A_{1}\mathrm{exp}\left(\frac{-\sigma^{2}t^{2}}{2}\right)\mathrm{cos}(w_{1}t+\phi)+A_{2}\mathrm{cos}(w_{2}t+\phi) .
\label{eqn1:Tranf}
\end{equation}
Here $w_{1}$ and $w_{2}$ are the frequencies of the muon precession signal and background signal respectively, $\phi$ is the initial phase offset, and $\sigma$ is the total depolarization rate. Asymmetry spectra were recorded at several temperatures above and below $T_{c}$, to determine the temperature dependence of depolarization rate $\sigma$. The background nuclear depolarization $\sigma_{\mathrm{N}}$ obtained from the spectra above T$_{c}$ was temperature independent over the temperature range of study with average value $\sigma_{\mathrm{N}}$ = 0.134 $\mu$s$^{-1}$. In order to obtain the depolarization due to superconducting state $\sigma_{\mathrm{FLL}}$, the background contribution was subtracted quadratically from the total sample depolarization rate $\sigma$ as per relation:

\begin{equation}
\sigma^{2} = \sigma_{\mathrm{N}}^{2}+\sigma_{\mathrm{FLL}}^{2} .
\label{eqn2:sigma}
\end{equation}

The resulting temperature dependence of muon-spin depolarization rate in the superconducting state $\sigma_{\mathrm{FLL}}$ is shown in Fig. 3(b). The depolarization rate increases systematically as the temperature is lowered below T$_{c}$ whereas the contribution above T$_{c}$ was fixed to zero. The $\sigma_{\mathrm{FLL}}$ is related to the London magnetic penetration depth $\lambda$ and thus to the superfluid density $n_{s}$ by the equation, 

\begin{equation}
\frac{\sigma_{\mathrm{FLL}}(T)}{\sigma_{\mathrm{FLL}}(0)} = \frac{\lambda^{-2}(T)}{\lambda^{-2}(0)} .
\label{eqn3:sfd}
\end{equation}

For a single-gap $\textit{s}$-wave superconductor in the dirty limit, the temperature dependence of the London magnetic penetration depth within the London approximation is given by 
\begin{equation}
\frac{\lambda^{-2}(T)}{\lambda^{-2}(0)} = \frac{\Delta(T)}{\Delta(0)}\mathrm{tanh}\left[\frac{\Delta(T)}{2k_{B}T}\right] ,
\label{eqn4:lpd}
\end{equation}
 where $\Delta$(T) = $\Delta_{0}$$\delta(T/T_{c})$. The temperature dependence of the gap in BCS approximation is given by the expression $\delta(T/T_{c})$ = tanh[1.82(1.018($\mathit{(T_{c}/T})$-1))$^{0.51}$]. The dirty limit model was used since Re$_{6}$X family crystallizes in the $\alpha$- Mn structure, which is already known to have an intrinsic disorder with high residual resistivity. This yields very low free path compared to BCS coherence length \cite{RPS,DAM,DSA,DSJ}.\\ 
We obtain good fits to the $\sigma_{\mathrm{FLL}}$(T) data using the model discussed above (see Fig. 3(b)). The fitted value for the transition temperature, T$_{c}$ = 5.98(2) K, is in good agreement with the value obtained (${T}_{c}$ = 6 K) from bulk measurements. The energy gap has a maximum magnitude of $\Delta$(0) = 0.95(2) meV, which yields the value $\Delta$(0)/k$_{B}$T$_{c}$ = 1.84(2) which is larger than the BCS expectation (1.764), indicating enhanced electron-phonon coupling strength. This behavior is in good agreement with our heat capacity measurements and common to the earlier studies on Re$_{6}$Hf and Re$_{6}$Zr where similar strong coupling limit was observed.

 The zero temperature effective magnetic penetration depth $\lambda$(0) can be directly calculated from the $\sigma_{\mathrm{FLL}}$(0) (0.4401(4) $\mu$s$^{-1}$) from the relation \cite{JES,EHB} 

\begin{equation}
\frac{\sigma_{\mathrm{FLL}}^2(0)}{\gamma_{\mu}^2} = 0.00371 \frac{\Phi_{0}^{2}}{\lambda^{4}(0)} ,
\label{eqn5:lam}
\end{equation} 

where $\gamma_{\mu}$/2$\pi$ = 135.53 MHz/T is the  muon gyromagnetic ratio and $\Phi_{0}$ is the magnetic flux quantum. The magnetic penetration depth at T = 0 K was thus found to be $\lambda$(0) = (4937 $\pm$ 11) \text{\AA}.

\begin{figure*}[t]
\includegraphics[width=2.0\columnwidth]{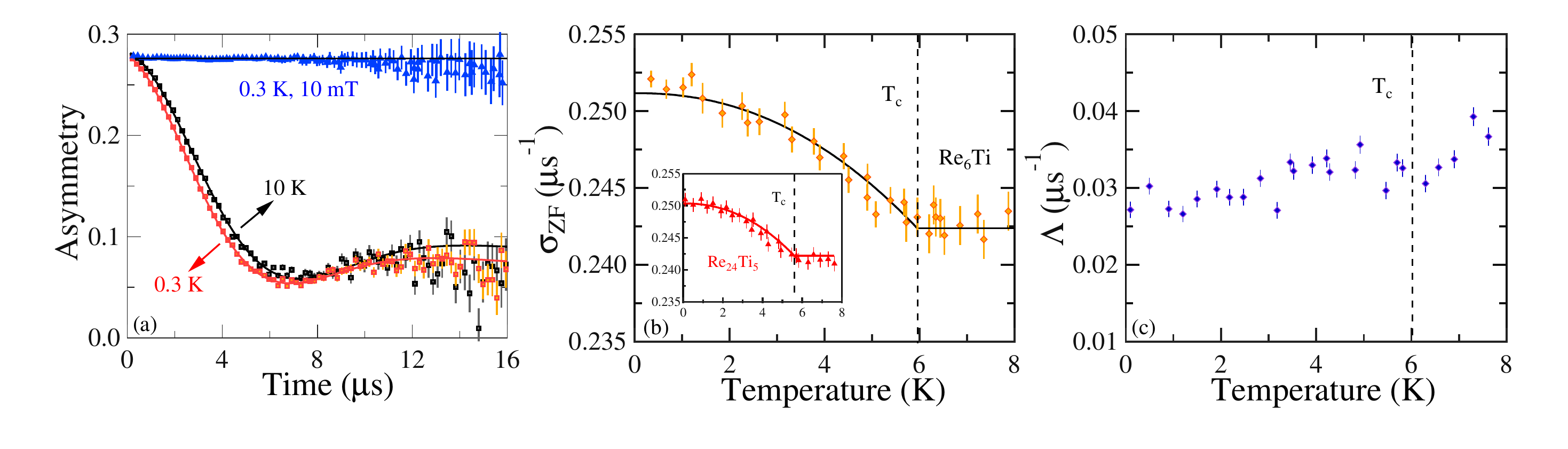}
\caption{\label{Fig4:TF} (color online)(a) Zero field $\mu$SR spectra collected below (0.3 K) and above (10 K) the superconducting transition temperature. The solid lines are the fits to Guassian Kubo-Toyabe (KT) function given in Eq. (7)  (b) Temperature dependence of nuclear relaxation rate $\sigma_{ZF}$shows systematic increase below $T_{c}$ (c) Temperature dependence of electronic relaxation rate $\Lambda$ shows no appreciable change at $T_{c}$.}
\end{figure*}

 Zero-field muon spin relaxation (ZF-$\mu$SR) measurements are carried out to detect the tiny spontaneous magnetic fields associated with the broken TRS in the superconducting state. The relaxation spectra was collected below ($T$ = 0.3 K) and above the transition temperature ($T_{c}$ = 10 K), as displayed in Fig. 4(a). There was no hint for an oscillatory component in the data, which suggests the absence of an ordered magnetic structure. Interestingly, the spectra traces a different relaxation channel below the transition temperature, which indicates the presence of the spontaneous internal magnetic field. In the absence of atomic moments the relaxation is due to randomly oriented nuclear moments, which can be modeled by the Gaussian Kubo-Toyabe (KT) function \cite{RSH}  
\begin{equation}
G_{\mathrm{KT}}(t) = \frac{1}{3}+\frac{2}{3}(1-\sigma^{2}_{\mathrm{ZF}}t^{2})\mathrm{exp}\left(\frac{-\sigma^{2}_{\mathrm{ZF}}t^{2}}{2}\right) ,
\label{eqn6:zf}
\end{equation} 
where $\sigma_{\mathrm{ZF}}$ accounts for the relaxation due to static, randomly oriented nuclear dipolar local fields at the muon site.
The zero field asymmetry spectra well described by the function
\begin{equation}
A(t) = A_{1}G_{\mathrm{KT}}(t)\mathrm{exp}(-\Lambda t)+A_{\mathrm{BG}} ,
\label{eqn7:tay}
\end{equation} 
where $A_{1}$ is the initial asymmetry, $\Lambda$ is the electronic relaxation rate, and $A_{\mathrm{BG}}$ is the time independent background contribution from the muons stopped in the sample holder. The above function was fitted to the ZF asymmetry spectra collected at various temperatures above and below T$_{c}$, which yields the temperature dependence of fit parameters $\sigma_{\mathrm{ZF}}$ and $\Lambda$ as shown in Fig 4(b) and 4(c). The sample and background asymmetries have approximately temperature independent values A$_{1}$ = 0.1732(4) and A$_{\mathrm{BG}}$ = 0.1158(4). Interestingly, the Gaussian relaxation rate parameter $\sigma_{\mathrm{ZF}}$ shows a clear increase below the temperature T = 5.98 $\pm$ 0.2 K [see Fig. 4(b)], which is close to superconducting transition temperature. Such a systematic increase in $\sigma_{\mathrm{ZF}}$ below T$_{c}$ was also identified in other members of Re$_{6}$X (X = Hf, Zr) \cite{RPS,DSJ} family by $\mu$SR measurements. This particular behavior was attributed to the formation of spontaneous magnetic fields below T$_{c}$, which in turn confirmed time-reversal symmetry breaking in this compounds. These observations clearly suggest that TRS is also broken in the superconducting state of Re$_{6}$Ti and the Re$_{6}$X is a unique family where all the members till now have shown this exotic phenomena.\\
\begin{table}[h!]
\caption{Comparison of mode of internal field calculated for Re$_{6}$X(X= Zr,Hf,Ti)}
\begin{center}
\begin{tabular}[b]{lc}\hline\hline
Compound & Internal Field\\
\hline                                  
Re$_{6}$Hf&0.085 \\
Re$_{6}$Zr&0.11\\                                    
Re$_{6}$Ti&0.14\\
\hline
Re$_{24}$Ti$_{5}$&0.13
\\[0.5ex]
\hline\hline
\end{tabular}
\par\medskip\footnotesize
\end{center}
\end{table}
 To eliminate the possibility that the above signal is due to extrinsic effects such as impurities, we applied 10 mT longitudinal field. As depicted by the blue markers in Fig. 4(a), this was sufficient to fully decouple the muons from the internal field, which appears as flat asymmetry spectra. This indicates that the associated magnetic fields are in fact static or quasistatic on the time scale of the muon precession. The magnitude of internal field can be calculated by:
\begin{equation}
\label{eq:8}
    |B_\mathrm{int}|=\sqrt{2}\frac{\sigma_{\mathrm{ZF}}}{\gamma_\mu},
\end{equation}
where $\gamma_\mu/2\pi=135.5~$MHz/T is the muon gyromagnetic ratio. The increase in $\sigma_{\mathrm{ZF}}$ below T$_{c}$ is $\sim$ 0.0084(1)~$\mu\mathrm{s}^{-1}$ which gives the internal field strength |B$_\mathrm{int}$| = 0.14 G. According to theoretical predictions, ASOC plays a pivotal role in mixing of spin-triplet/spin-singlet pairing. Since spin-orbit coupling varies as Z$^{4}$, it is expected that its strength would be weaker in Ti as compared to Hf and Zr. If this is the case, then it should reduce the spin singlet/triplet pairing mixing ratio whose direct consequence must be visible in ZF-$\mu$SR. In contrast, we observed similar results where the TRSB signal observed in Re$_{6}$Ti is remarkably identical in magnitude to that seen in other members of Re$_{6}$X family. In addition, the value of the calculated internal field is also comparable to the values obtained for Re$_{6}$Hf and  Re$_{6}$Zr as shown in Table-I. This suggests that the effect of spin-orbit coupling does not lead to an increase in the strength of the spin-triplet channel in these compounds. 

 Recent reports have shown that the Re density of states (DOS) near Fermi level is the major contributor to the total DOS in the Re$_{6}$X family \cite{MAK}. It was hinted that these could be the reason for the negligible effect of spin-orbit coupling and repeated occurrence of TRSB in all these compounds. To confirm that, we have done the zero-field $\mu$SR measurements in the isostructural and stoichiometric compound Re$_{24}$Ti$_{5}$ with $T_{c}$ =5.87 K, where we have comparatively reduced the Re composition (ratio of 4.8:1) such that the possible influence of Re at the density of states can be reduced. Interestingly, the ZF-$\mu$SR in this compound also shows TRSB, with similar nature and magnitude of the temperature dependence of Gaussian relaxation rate $\sigma_{\mathrm{ZF}}$ [see inset Fig. 4(b)]. Also, the internal field |B$_\mathrm{int}$| have approximately the same magnitude as seen for other Re$_{6}$X compounds. It indicates that in Re$_{6}$X family of compounds mixing of spin-singlet and triplet channel may be controlled by some other mechanism and need a detailed theoretical work.

 In conclusion, we have determined that the superconducting ground state in Re$_{6}$Ti and Re$_{24}$Ti$_{5}$ breaks time-reversal symmetry, which is in addition to other members of Re$_{6}$X (X=Hf, Zr) family which shows these phenomena. The TF data suggest that the superconducting order parameter is described well by an isotropic gap with s-wave pairing symmetry with enhanced electron-phonon coupling, similar to that of Re$_{6}$Hf and Re$_{6}$Zr. The emergence of similar results suggests that that the strength of spin-orbit coupling does not affect the pairing symmetry in Re$_{6}$X family compounds. Further theoretical and experimental work required to understand the unconventional superconductivity of the Re$_{6}$X family of compounds.
\section{Acknowledgments}
R.~P.~S.\ acknowledges Science and Engineering Research Board, Government of India for the Young Scientist Grant YSS/2015/001799 and Ramanujan Fellowship through Grant No. SR/S2/RJN-83/2012. We thank ISIS, STFC, UK for the muon beamtime and Newton Bhabha funding to conduct the $\mu$SR experiments.

\end{document}